# Toward a Collection-based Metadata Maintenance Model

Martin Kurth
Cornell University Library
Tel: +1 607 255 6112
Fax: +1 607 255 6110
MK168@cornell.edu

Jim LeBlanc
Cornell University Library
Tel: +1 607 254 5290
Fax: +1 607 255 6110
JDL8@cornell.edu

**Abstract**:
In this paper, the authors identify key entities and relationships in the operational management of metadata catalogs that describe digital collections, and they draft a data model to support the administration of metadata maintenance for collections. Further, they consider this proposed model in light of other data schemes to which it relates and discuss the implications of the model for library metadata maintenance operations.

## 1. Introduction

In a recent essay, Ruth Bogan, Head of Database and Catalog Portal Management at the Rutgers University Libraries, articulates a key issue: "All library technical services managers face an essential question: How does one direct a workforce toward a future that cannot be seen?"(1) Bogan goes on to observe that the expertise and skills which have long been the hallmark for the maintenance of libraries' catalog data can and must be parlayed towards metadata management in a broader set of information delivery systems. Although Bogan focuses chiefly on the "manual," record-by-record components of library data maintenance, her call to expand the scope of this traditional function to include metadata schemes other than MARC is an important one. If nothing else, non-MARC metadata for library digital collections is often propped on existing MARC metadata and, if the latter is routinely maintained through traditional database management workflows (either manual or automated), corresponding paradigms must be in place for the upkeep of the non-MARC data that is derived from the information stored in the library's catalog.(2)

One of the more significant obstacles to creating such paradigms lies in the fact that administrators of the metadata for individual collections are often scattered throughout the library environment, usually outside the technical services or systems departments charged with the initial creation of a collection's metadata. Responsibility for maintaining this data, or even the idea that this information may need to be continually maintained, can fall between organizational and administrative cracks. The growing use of digital collection registries to collocate descriptive, administrative, and technical metadata about digital collections represents a significant step toward building a framework for closing or bridging these cracks. However, most collection registry models tend to focus primarily on the content objects gathered into collections and only secondarily on the metadata catalogs that describe those collections.(3) More work needs to be done to develop data models to support operations that maintain, update, and correct digital collection metadata catalogs over time.

This paper describes key entities and relationships in the ongoing operational management of digital collection metadata catalogs and drafts for discussion a data model that would support the administration of programmatic approaches to metadata maintenance operations. By proposing this model, we hope to engage the communities working with digital collection descriptions to resolve any inconsistencies between the present model and existing ones, to normalize elements and vocabularies that can be shared across models, and to determine productive strategies for interleaving metadata maintenance description sets with collection description sets.

The remainder of the paper falls into four sections. Section 2 offers an operational context for the metadata maintenance model we are proposing. Section 3, the heart of the paper, describes the model itself. In Section 4 we relate the model to other data models important to it. Finally, we conclude in Section 5 by recapping the significant features of the model and discussing its implications for metadata maintenance operations.

## 2. Operational context

As a first step in modeling metadata maintenance operations, we offer a preliminary list of ten metadata maintenance functions that are typical when administering metadata records for content objects in collections. These maintenance functions are:

- Accrual (or addition of new records)
- Deletion
- Modification
- Transformation
- Reporting
- Export
- Mapping
- Migration
- Exposure
- Activation / deactivation

All of the maintenance functions may conceivably come into play as the nature and content of a digital object or collection change. Librarians and library programmers already know how to perform these functions for given targets, though it is not always clear how the various practitioners of this work must interact in the broader sphere of interrelated objects and collections. The elements and values that underpin these interactive relationships must be identified, defined, and codified in order to insure the efficient functioning of the information system and the ongoing accuracy and integrity of the system's data.

In his now classic schema, J.A. Zachman presents a descriptive framework for systems architecture that is of particular value for businesses and institutions in which technology and effort are distributed.(4) The "Zachman Framework" describes entities and relationships within a given system in terms of six generic attributes: what, how, where, who, when, and why. By identifying and associating elements in the system in this way, Zachman is able to construct a multidimensional description of interrelationships among work teams and the tasks and/or products they deliver. Underlying this description is an understanding that individual pieces of the overall framework must be tailored to specific stakeholder perspectives. Zachman uses an architectural example to illustrate how the values inherent in the owner's, the designer's, and the builder's points of view may differ with regard to a structure. Elements of the Zachman framework may thus vary in nature, terminology, and level of detail, depending on the stakeholders at whom the particular elements are aimed.

The hexagonal diagram in Figure 1 on the following page illustrates how a Zachman-type model can be applied to a metadata maintenance situation.(5) The six rectangular boxes represent the pronominal attributes associated with metadata maintenance for a collection. In addition to the maintenance function itself, the attributes include those that typically exist in relation to each maintenance function: Periodicity, the frequency at which administrators should perform the function; Documentation that describes automated and/or manual workflows associated with the function; Scripts or Services, the programming tools that engage the maintenance function; the administrative Department responsible for performing the function; and Contact, the individual or group designated to receive communications regarding the function. Although Department and Contact may be redundant for maintenance functions carried out in a small operation, these entities may differ in a larger, more distributed environment. For example, in the latter case the Contact may be a collection's administrator, while the Department may be the work unit where the metadata maintenance is actually performed. The linearly defined facets of the diagram reveal the interrelationships among each attribute for each maintenance function. The real world context for this description is much more complicated, of course, and one must imagine ten levels (representing the ten metadata maintenance functions proposed above), with interrelational linkage in three dimensions among the individual boxes and hexagons at all levels, to visualize the complete framework on which data maintenance for a given collection should ideally be managed.

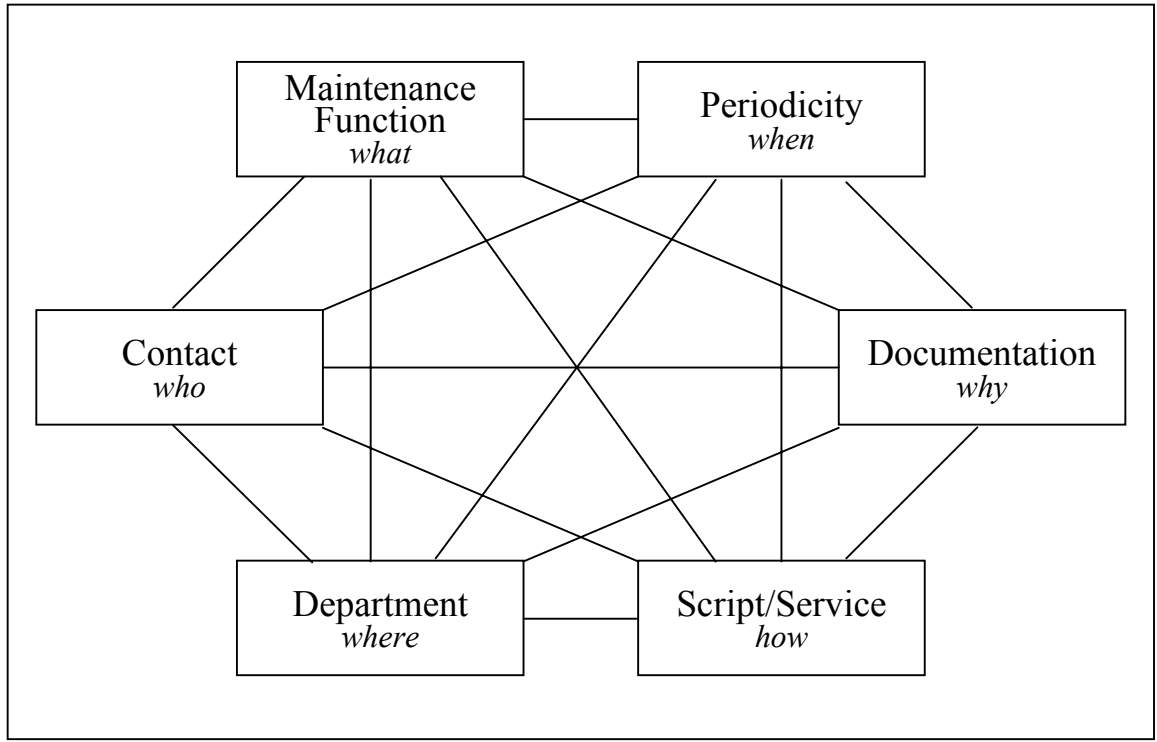

*Figure 1.    Zachman-type diagram for metadata maintenance operations*

Given this operational view of library data maintenance, what kind of metadata model, existing or new, is best suited for the administrative and technical management of these functions? In Section 3, we outline the significant features of a metadata model to support the administration of metadata maintenance for collections.

## 3.  The metadata maintenance model

In *An Analytical Model of Collections and Their Catalogues*, Michael Heaney describes the recursive scenario in which a metadata catalog, or analytical finding-aid, describes a collection and is in turn described by a unitary finding-aid.(6) Figure 2 depicts these relationships in the context of metadata maintenance, using relevant elements and vocabulary terms from the Dublin Core Collection Description Application Profile (DC CD AP) and the Collection Type (CLDType) Vocabulary.(7) In Figure 2, Collection A is a collection of content objects, as reflected by it having one or more CLDTypes that designate it in terms of the content of the items within it, such as cldtype:CollectionImage for a collection of images and cldtype:CollectionPhysicalObject for a collection of physical objects. Collection A is described by one or more collections (Collection B, etc.) that are catalogs of metadata records, as reflected by the cld:collectionDescription attribute that relates the metadata catalogs to Collection A and by their CLDType designation of Catalogue. As we have explained, it is the ongoing maintenance of such metadata catalogs that interests us here. To that end, we represent in the figure that one or more collection descriptions describe the metadata catalogs for the purposes of metadata maintenance, as reflected by those collection descriptions' adherence to a yet-undeveloped Metadata Maintenance Application Profile (MDM AP). We begin to sketch out the rudimentary features of such a metadata maintenance collection description in the remainder of this section.

As noted, Figure 2 allows for the existence of one or more collection descriptions that describe a given metadata catalog. This illustrates our observation that maintenance administrators of metadata catalogs are often scattered throughout an organization or even among collaborating organizations. Such distributed environments may require multiple descriptions to represent key components of the operations of multiple units, so we do not prescribe that such representations be folded into a single description. We leave those considerations to implementers of this model.

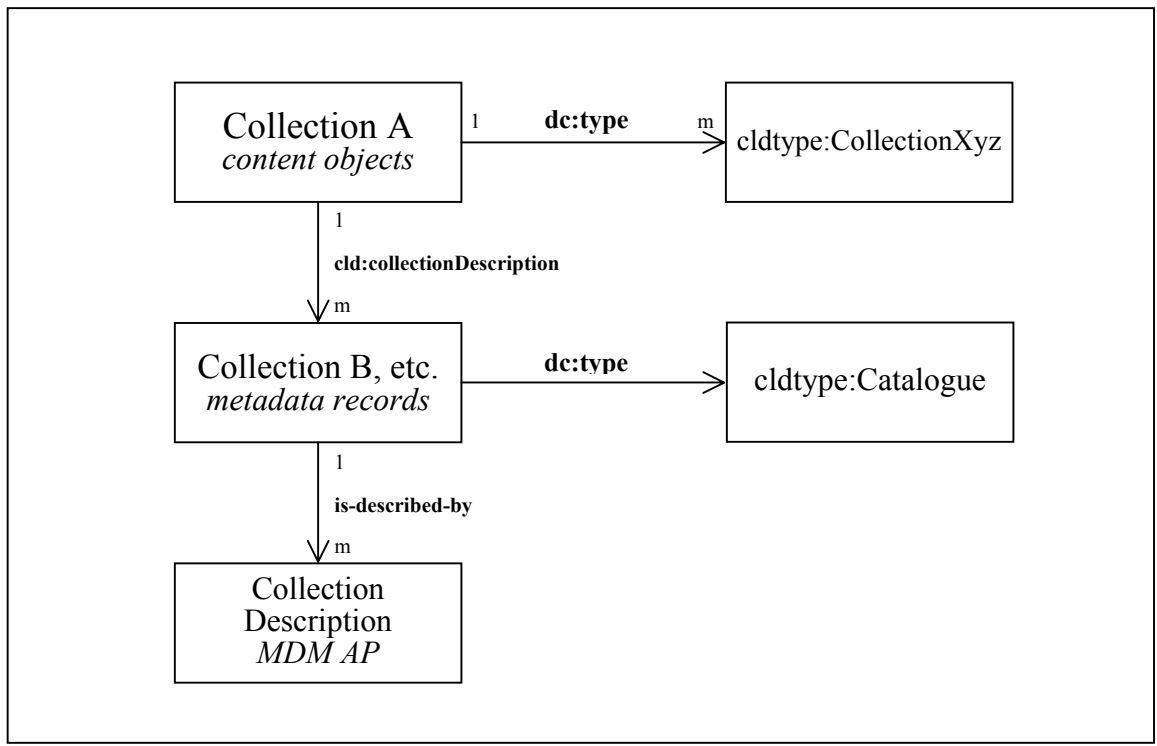

*Figure 2. Metadata maintenance collection descriptions in relation to the metadata catalogs they describe*

To summarize the important relationships in Figure 2: A metadata maintenance collection description (adhering to a proposed MDM AP) describes a metadata catalog (Collection B), which in turn describes a collection of content objects (Collection A). An important question regarding these relationships remains, though it is beyond the scope of this paper: Should a metadata maintenance collection description be represented in a collection-level record for the collection of content objects to which it relates (such as Collection A in Figure 2)? In other words, should collection description application profiles, such as those adhering to the DC CD AP, include a "pointer" to any metadata maintenance descriptions that exist in relation to collections of content objects? Such decisions are likely to be primarily local implementation decisions, but some collective thinking about the issue would benefit metadata maintenance administrators.

With the relationships in Figure 2 thus established, we use Figure 3 to depict in detail the properties and resources that we expect would combine to form a metadata maintenance collection description. The model primarily comprises properties that are attributes of a metadata catalog, that is, a Collection of cldtype:Catalogue, and properties that are attributes of one or more Maintenance Functions that operate on the metadata catalog.

With regard to the properties that are attributes of a metadata catalog, we first propose that the catalog has a dc:type property that would adhere to an mdm:MDMCollType encoding scheme. An MDMCollType vocabulary would enable metadata maintenance administrators to declare that metadata catalogs were of types such as Legacy (metadata records inherited from another source), Storage ("canonical" metadata records used to derive records for various purposes), or Delivery (metadata records used in a particular delivery system).

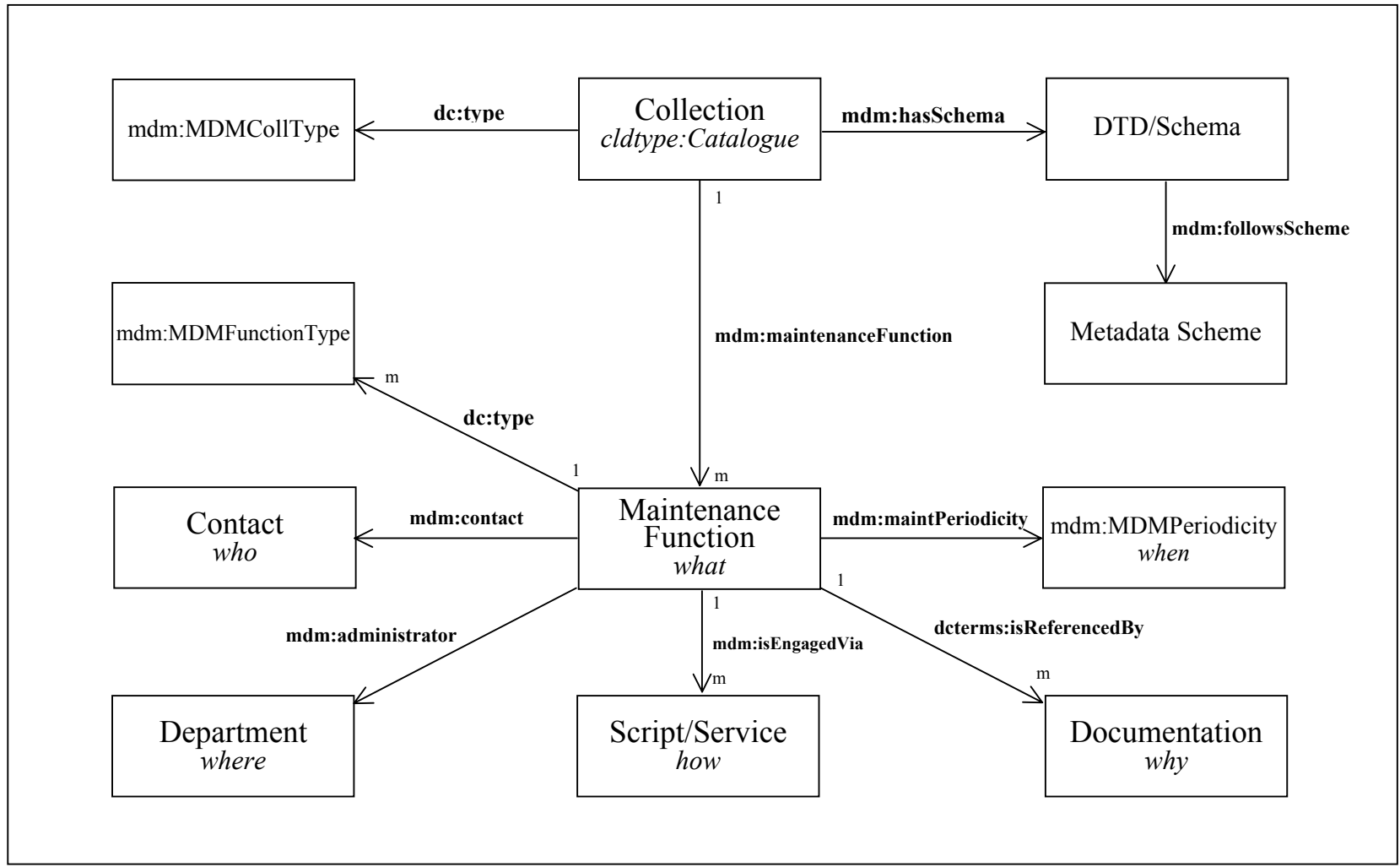

*Figure 3. Detailed representation of the properties and resources comprising a metadata maintenance collection description*

Next we declare that the metadata catalog is validated by a particular locally-modified XML document type definition (DTD) or schema, that is, that it has an mdm:hasSchema property whose value would typically be a uniform resource identifier (URI) for the DTD or schema.(8) Representing applicable DTDs and schemas as resources in this way would allow maintenance administrators to identify all metadata catalogs adhering to a particular DTD or schema. Using similar logic, we indicate that the DTD or schema follows a particular metadata scheme, such as the Dublin Core Metadata Element Set, by declaring that the DTD or schema has an mdm:followsScheme property whose value would be the name or identifier of the scheme. The model provisionally provides for mdm:hasSchema and mdm:followsScheme properties, but in fact these relationships call for further research to determine whether properties from existing schemas would be sufficient to meet the needs identified here.

Finally, with regard to metadata catalog properties, central in the figure is the relationship between a metadata catalog and the one or more Maintenance Functions that operate on it. Each Maintenance Function is an instance of one or more metadata maintenance function types, whose values are controlled by a yet-to-be-defined MDMFunctionType vocabulary that would reside in an "mdm" metadata maintenance namespace. Candidate values for the mdm:MDMFunctionType scheme are the names of the ten metadata maintenance functions listed in Section 2. It may be useful for readers to see the mdm:maintenanceFunction property that relates a metadata catalog (Collection of cldtype:Catalogue) to a Maintenance Function (instance of one or more MDMFunctionTypes) roughly as an expansion of the dcterms:accrualMethod property that relates a Collection to a cld:DCCDAccrualMethod value in the DC CD AP.(9) The difference lies in that whereas the DC CD AP is solely interested in accrual as a means of modifying a collection, our metadata maintenance model is interested in collection modification more broadly and has thus proposed mdm:maintenanceFunction as a generic property that relates collections to instances of all types of maintenance operations, as enumerated by the ten maintenance functions proposed in Section 2.

In Figure 3, a Maintenance Function instance serves in turn as a resource with properties of its own, thereby collocating the other five metadata maintenance entities diagrammed in Figure 1: Periodicity, Documentation, Script/Service, Department, and Contact. We believe that rendering these entities as properties of Maintenance Function is justified in light of the catalog maintenance operations with which we are familiar; that is, administrators typically regard staffing, procedures, tools, and task frequency in terms of the maintenance functions they support. Further, grouping Periodicity, Documentation, Script/Service, Department, and Contact by Maintenance Function serves as a de facto categorization of these entities, thus simplifying the model by reducing the need to define additional properties that describe them.

To render Periodicity in Figure 3, we base our periodicity-related property (mdm:maintPeriodicity) and encoding-scheme-controlled value (mdm:MDMPeriodicity) on the parallel periodicity-related property (dcterms:accrualPeriodicity) and encoding scheme (cld:DCCDAccrual Periodicity) defined in the DC CD AP.(10) Whereas we feel that differentiating a "maintenance periodicity" property from Accrual Periodicity is warranted because Accrual Periodicity is defined specifically as "the frequency with which items are added to a collection," we question the necessity of defining a specific "metadata maintenance periodicity" encoding scheme for the various frequencies of metadata maintenance operations. Currently the DCCDAccrualPeriodicity encoding scheme is defined as applying to "frequencies with which items are added to a collection."(11) Renaming

**Properties of a collection of content objects**

| | |
|---|---|
| dc:type | Has CLDType value that categorizes the collection according to the content of the items that make it up, e.g., cldtype:CollectionImage. |
| cld:collectionDescription | Has value that identifies a collection (metadata catalog) of cldtype:Catalogue. |

**Properties of a metadata catalog**

| | |
|---|---|
| dc:type | Has value of cldtype:Catalogue |
| dc:type | Has value from mdm:MDMCollType encoding scheme, e.g., Legacy, Storage, Delivery. |
| mdm:hasSchema | Has value that identifies the DTD or schema that validates records in the catalog. |
| mdm:maintenanceFunction | Has value that identifies a maintenance function that operates on the catalog. |

**Properties of an XML DTD or schema**

| | |
|---|---|
| mdm:followsScheme | Has value that identifies the metadata scheme to which the DTD or schema adheres. |

**Properties of a metadata maintenance function**

| | |
|---|---|
| mdm:maintPeriodicity | Has value from mdm:MDMPeriodicity encoding scheme. |
| dcterms:isReferencedBy | Has value that identifies a Documentation instance that describes automated or manual processes related to the maintenance function. |
| mdm:isEngagedVia | Has value that identifies a Script or Service that engages the function. |
| mdm:administrator | Has value that identifies a Department responsible for the function. |
| mdm:contact | Has value that identifies the point of contact (e.g., email address or web form) for an individual or group to whom communication regarding the function should be directed. |
| dc:type | Has value from mdm:MDMFunctionType encoding scheme, e.g., Accrual, Deletion, Modification. |

*Figure 4.   Properties of the primary entities in a metadata maintenance collection description*

DCCDAccrualPeriodicity and defining it more broadly would enable its use with additional properties, such as the mdm:maintPeriodicity property we are proposing here. Lacking such a broadened Periodicity encoding scheme for more general use, we elected provisionally to propose an MDMPeriodicity encoding scheme in Figure 3.

The rendering of the remaining metadata maintenance entities of Documentation, Script/Service, Department, and Contact in Figure 3 should largely be self-explanatory. One or more Documentation instances relate to a Maintenance Function via the dcterms:isReferencedby property. We propose an mdm:isEngagedVia property to relate one or more Scripts or Services to a Maintenance Function. And, finally, we propose mdm:administrator and mdm:contact properties to relate a Department and Contact to a Maintenance Function, respectively.(12)

The contents of this section and Figures 2 and 3 outline the significant features of a metadata maintenance collection description. Figure 4 summarizes the properties of the primary entities involved. Clearly, more work needs to be done to define properties, value types, and encoding schemes, and to develop the model described here into a Metadata Maintenance Application Profile. By way of providing a context for some of this remaining work, we discuss the present model in the following section in light of other important data models to which it relates.

## 4. Relationship to other data models

The discussion of the features of a metadata maintenance model in Section 3 has brought to light the ways in which it relates to Heaney's *Analytical Model of Collections and Their Catalogues*, the Dublin Core Collection Description Application Profile, and (in a note) the Research Support Libraries Programme Collection Description Schema. Continuing development of the DC CD AP and its related vocabulary encoding schemes in particular should provide opportunities for dialog with DC CD AP developers regarding points of intersection with a metadata maintenance collection description model. Moreover, further development of a metadata maintenance model also warrants study of other data models for design features that can influence that development.

In one such model, Christophe Blanchi and Jason Petrone describe an object-based architecture for metadata management.(13) Though their approach is object-based and the approach described here is collection-based, both models use the strategy of declaring metadata schema and metadata management services as high-level objects in order to manage their relationships to the metadata to which they obtain. In addition, the Blanchi and Petrone model applies rigorous typing and identification strategies to all resources as a foundation upon which to build automated metadata management services. These strategies and the services built on them bear further analysis for their applicability to collection-based metadata maintenance services.

In another data model, the one underpinning the Global Digital Format Registry (GDFR), Stephen L. Abrams accounts for maintenance services that involve "creation, updating, and deletion" of entries in the registry.(14) The GDFR model defines Maintenance as a high-level property of the registry and designates Maintenance as an exemplar of the Authority data type, which captures information regarding the agency responsible for format maintenance.(15) More importantly, GDFR developers have designed its data model to subtend "service gateways" that enable both human and automated processes, which directly parallels the mix of services that the metadata maintenance model intends to support.(16) The GDFR's attention to interrelated data and service models designed to support format maintenance calls for close study of the GDFR as its developers continue to refine it.

When discussing the high-level properties of the GDFR, Abrams observes that maintenance agencies are "associated with specific, though possibly unbounded, time spans."(17) Indeed, the metadata maintenance model described here as yet takes no account of the temporal attributes of the properties of metadata catalogs and maintenance functions. That temporality is inherent in the resources associated with maintenance operations justifies analyzing not only how the GDFR data model renders time dependencies but also how other models that feature temporality, such as the ABC Model, render them as well.(18)

To sum up, further refinement of a metadata maintenance model requires more rigorous testing of its data structures in light of their relationships to other data models that address collection descriptions, metadata management, metadata maintenance processes, and the temporality of maintenance resources. In the final section of this paper, we highlight the significant features of the metadata maintenance collection description model and discuss its potential benefits for metadata maintenance operations.

## 5. Conclusion

The metadata model described here offers a simple scheme for organizing the resources involved in metadata maintenance operations, such as documentation, scripts, and contacts. Further, it provides a structure for storing and retrieving metadata maintenance information as it relates to specific collections of content objects, metadata catalogs, DTDs, schemas, metadata schemes, and so on. The model might also be adapted for maintenance of collections of content objects.

We have sought simplicity in the model in order to keep down the costs involved in gathering, storing, and managing catalog maintenance metadata. Moreover, we have been mindful of sustainability issues in this early iteration of the model and are particularly concerned that they continue to inform its further development. Though failing to track maintenance data involves implicit costs to an organization, especially as collections (and the staff managing them!) age, we anticipate that the need for yet another cache of metadata (albeit in support of a worthy cause) is likely to be met with skepticism by fiscally accountable administrators. Therefore we expect that refinement of this model will also involve articulating clear use and business cases for its implementation.

It is in this regard, however, that library technical services managers may choose to leverage both the traditional skills of their catalog management workforce and the potential applications of the metadata maintenance metadata scheme described above to address the fundamental operational management questions posed by Zachman for any complex or distributed workforce. Even when faced with limited resources for ongoing metadata upkeep, the key elements of these operational and metadata maintenance models can provide a rigorous basis for developing and discussing workflow options and for setting data maintenance priorities at an institutional or multi-institutional level.

## References

1. Ruth A. Bogan. Redesign of Database Management at Rutgers University Libraries. In *Innovative Redesign and Reorganization of Library Technical Services: Paths for the Future and Case Studies*. Westport, CT: Libraries Unlimited, 2004, p. 161.
2. For an examination of some of the issues inherent in repurposing MARC data for non-MARC applications, see: Martin Kurth, David Ruddy, and Nathan Rupp. Repurposing MARC Metadata: Using Digital Project Experience to Develop a Metadata Management Design. *Library Hi Tech*, 2004, v. 22, n. 2, pp. 153-165.
3. The following approaches to collection descriptions represent metadata catalogs within their data models. They define a metadata catalog, also known as an analytic finding aid, as a collection of metadata records that describes the items (and the intellectual content of the items) in a collection of content objects. Though both approaches include metadata catalogs in their models, they nevertheless emphasize collection descriptions as tools for discovering and administering collections of content objects, rather than as tools for administering collections of metadata records: Heaney, Michael. *An Analytical Model of Collections and their Catalogues*. 3rd issue, rev. UKOLN/OCLC, January 2000. http://www.ukoln.ac.uk/metadata/rslp/model/amcc-v31.pdf; *Dublin Core Collection Description Application Profile*. 2006-02-24. http://www.ukoln.ac.uk/metadata/dcmi/collection-application-profile/.
4. J.A. Zachman. A Framework for Information Systems Architecture. *IBM Systems Journal*, 1987, v. 26, n. 3, pp. 276-292.
5. This diagram is adapted from the hexagonal model found in: J.F. Sowa and J.A. Zachman. Extending and Formalizing the Framework for Information Systems Architecture. *IBM Systems Journal*, 1992, v. 31, n. 3, p. 611.
6. Heaney. *An Analytical Model of Collections and their Catalogues*, pp. 4-5.
7. *Dublin Core Collection Description Application Profile*; *Collection Type (CLDType) Vocabulary*. 2005-3-19. http://www.ukoln.ac.uk/metadata/dcmi/collection-type/.
8. The model as expressed in Figure 3 currently assumes that the metadata catalogs to be maintained are XML-based. We recognize that we would need to modify or expand the model in order to address non-XML metadata catalogs such as Resource Description Framework (RDF) triple stores.
9. Accrual Method. In *Dublin Core Collection Description Application Profile*. http://www.ukoln.ac.uk/metadata/dcmi/collection-application-profile/#dctermsaccrualmethod.
10. Accrual Periodicity. In *Dublin Core Collection Description Application Profile*. http://www.ukoln.ac.uk/metadata/dcmi/collection-application-profile/#dctermsaccrualperiodicity.

11. *Dublin Core Collection Description Proposed Term : DCCDAccrualPeriodicity*. 2004-08-18. http://www.ukoln.ac.uk/metadata/dcmi/collection-DCCDAccrualPeriodicity/.
12. We considered co-opting the Research Support Libraries Programme (RSLP) Collection Description Schema's rslpcd:administrator property, defined as the "agent who has responsibility for the physical or electronic environment in which the collection is held," but the RSLP definition of the property differed sufficiently from our intended use to warrant proposing our own mdm:administrator property. For more information about the RSLP Collection Description Schema, see: Pete Johnston. *The RSLP Collection Description Schema*. Collection Description Focus Briefing paper 2. UKOLN, September 2003. http://www.ukoln.ac.uk/cd-focus/briefings/bp2/; *RSLP Collection Description*. http://www.ukoln.ac.uk/metadata/rslp/.
13. Christophe Blanchi and Jason Petrone. Distributed Interoperable Metadata Registry. *D-Lib Magazine*, 2001, v. 7, n. 12. http://www.dlib.org/dlib/december01/blanchi/12blanchi.html.
14. Stephen L. Abrams. Establishing a Global Digital Format Registry. *Library Trends*, 2005, v. 54, n. 1, p. 139.
15. Abrams. Establishing a Global Digital Format Registry, p. 136; *Global Digital Format Registry (GDFR) Data Model*, v.4. 2004-01-12. http://hul.harvard.edu/gdfr/documents/DataModel-v4-2004-01-12.doc.
16. Abrams. Establishing a Global Digital Format Registry, p. 140.
17. Abrams. Establishing a Global Digital Format Registry, p. 136.
18. Carl Lagoze and Jane Hunter. The ABC Ontology and Model. *Journal of Digital Information*, 2001, v. 2, n. 2. http://jodi.ecs.soton.ac.uk/Articles/v02/i02/Lagoze/.